\documentclass{article}
\PassOptionsToPackage{numbers,compress}{natbib}
\usepackage[preprint]{neurips_2026} 
\workshoptitle{Beyond Private Training: The New Landscape of AI Privacy (InfPriv)}

\providecommand{\linenumbers}{}
\providecommand{\nolinenumbers}{}

\usepackage[utf8]{inputenc}
\usepackage[T1]{fontenc}
\usepackage{hyperref}
\usepackage{url}
\usepackage{booktabs}
\usepackage{amsfonts}
\usepackage{nicefrac}
\usepackage{microtype}
\usepackage{graphicx}
\usepackage{xcolor}
\usepackage{enumitem}
\usepackage{tabularx}
\usepackage{array}
\usepackage{csquotes}
\usepackage{xspace}

\usepackage{mdframed}
\usepackage{amssymb}

\newcommand{\dsname}{\textbf{\texttt{PIA-Bench}}\xspace}

\usepackage[breakable,listings,skins]{tcolorbox}
\usepackage{listings}
\definecolor{promptborder}{HTML}{A7A7A7}
\definecolor{promptbackground}{HTML}{FCFCFC}
\definecolor{promptplaceholder}{HTML}{C94F32}
\lstdefinestyle{piaprompt}{
  basicstyle=\ttfamily\fontsize{7.4}{9.0}\selectfont,
  columns=fullflexible,
  keepspaces=true,
  showstringspaces=false,
  breaklines=true,
  breakatwhitespace=true,
  tabsize=2,
  upquote=true,
  literate={—}{{\textemdash}}1,
  escapeinside={(*@}{@*)}
}
\newtcblisting{promptbox}{
  enhanced,
  breakable,
  listing only,
  listing engine=listings,
  listing options={style=piaprompt},
  colback=promptbackground,
  colframe=promptborder,
  boxrule=0.45pt,
  arc=0pt,
  outer arc=0pt,
  left=6pt,
  right=6pt,
  top=6pt,
  bottom=6pt,
  boxsep=0pt,
  before={\par\addvspace{6pt}\nolinenumbers},
  after={\par\addvspace{7pt}\linenumbers}
}
\newcommand{\promptvar}[1]{\textcolor{promptplaceholder}{\ttfamily\{#1\}}}

\title{\dsname: Towards Automated Privacy Impact Assessment with Large Language Models }

\author{%
  Jiamin Zheng \\
  School of Informatics \\
  University of Edinburgh \\
  Edinburgh, United Kingdom \\
  \texttt{jiamin.zheng@ed.ac.uk} \\
  \And
  Hao-Ping (Hank) Lee \\
  Carnegie Mellon University \\
  Pittsburgh, PA, United States \\
  \texttt{haopingl@cs.cmu.edu} \\
  \AND
  Luo Mai \\
  School of Informatics \\
  University of Edinburgh \\
  Edinburgh, United Kingdom \\
  \texttt{luo.mai@ed.ac.uk} \\
  \And
  Jingjie Li \\
  School of Informatics \\
  University of Edinburgh \\
  Edinburgh, United Kingdom \\
  \texttt{jingjie.li@ed.ac.uk}
}

\begin{document}

\maketitle

\begin{abstract}
Privacy impact assessment (PIA) is a critical instrument for institutions to proactively identify privacy risks and develop mitigation strategies before system deployment. 
While mandated across regulatory and institutional contexts, executing PIA requires extensive privacy and technical expertise, posing a particular challenge for teams without access to such resources. Prior work shows the potential of leveraging large language models (LLMs) to assist practitioners' privacy decisions, but little is known about how accurately and reliably LLMs can automate PIA. To this end, we develop \dsname, the first open benchmark for evaluating LLMs on real-world PIAs. We first audited 499 expert-authored PIAs published by US federal agencies and curated 73 structured PIAs, comprising a total of 451 privacy risk and 831 mitigation items, to evaluate LLMs' ability to assess privacy risks and propose mitigations of complex systems. Our results show that off-the-shelf LLMs produce meaningful assessments and identify avenues for future improvement. Finally, we call for improving domain-specific workflows for LLM agents, developing accountable LLM infrastructure, and designing new quality standards for PIAs.


\end{abstract}

\section{Introduction}
\label{sec:intro}
Privacy impact assessments (PIAs) have been a standardized approach for institutions to assess and document the privacy risks  when acquiring or designing digital technologies~\citep{wright2012}. PIAs help teams operationalize best practices in privacy engineering through modeling threats, designing proactive controls, and integrating privacy principles early in the development process rather than as an afterthought~\citep{iwaya2024, wright2012}. PIAs are also enforced in multiple regulatory contexts. US federal agencies are required to conduct and publish PIAs for systems that collect personal information \citep{egovact2002, omb2003}, while EU's General Data Protection Regulation (GDPR) mandates a parallel data protection impact assessment \citep{gdpr2016, wp29-2017}. 
However, preparing a PIA requires strong privacy expertise and is labour-intensive, which are resources that not every team possess~\citep{lee2024idont}. The complicated data practices in today's digital systems often exceed what any one person or team can meaningfully assess unaided or keeping up with continual updates~\citep{sadeh2026,lee2024idont, nicenboim2022, iwaya2024}. 

Large language models (LLMs) demonstrate promising capabilities in privacy reasoning and detecting privacy violations with little to no training effort~\cite{mireshghallah2024confaide}. Nevertheless, existing privacy benchmarks for LLMs primarily assess models' privacy \textit{behaviors}, for example, resistance to attacks that reveal private information~\citep{li2024privlm} or the ability to make contextually appropriate information disclosures~\citep{mireshghallah2024confaide}. More directly relevant to PIA, adjacent work has applied LLMs to broader risk and compliance analysis, such as identifying appropriate use cases for AI technologies under the EU AI Act~\citep{herdel2024exploregen} or supporting prototypers in interactively identifying and refining potential safety harms~\citep{wang2024farsight}. Privy further leverages LLMs to help AI practitioners identify privacy risks and curate mitigation plans~\citep{lee2026privy}. However, prior work either focuses on deployment- or runtime-stage use of LLMs rather than the complex, early-design-time context that PIA requires, or depends on heavy human interaction.

This gap motivates our work to explore \textbf{how to enable LLMs to automate the end-to-end PIA process at scale and responsibly}, transforming today's paradigm that is costly. We present \dsname, the first open-source benchmark, for evaluating LLMs' performance in real-world PIAs as the foundation for this vision. We collected and audited 499 U.S. federal PIAs, identified barriers to automation and selected 73 PIAs with sufficient structural contexts, yielding 451 passages for privacy risk and 831 for risk mitigation for real-world PIA benchmarking. By evaluating five popular LLMs, we identify key opportunities for future work, including the incorporating PIA-specific knowledge into structured workflows, supports truth-worthy human--LLMs collaborations, and develop broader quality metrics. We release \dsname in a public repository upon publication.



\label{sec:corpus}

\section{\dsname}
\label{sec:curation}
We created \dsname to evaluate LLMs' ability to identify privacy risks and propose mitigations from real-world system contexts. The benchmark contains 73 structurally separable PIAs from four U.S. federal agencies. For each PIA, we standardized the system context and extracted the agency's documented privacy risks and mitigations as ground truth. We evaluate coverage and specificity against expert ground truth and conduct error analysis to identify opportunities for improvement. Below, we detail the \dsname development and evaluation.

\begin{figure}[htbp]
    \centering
    \includegraphics[width=0.75\linewidth]{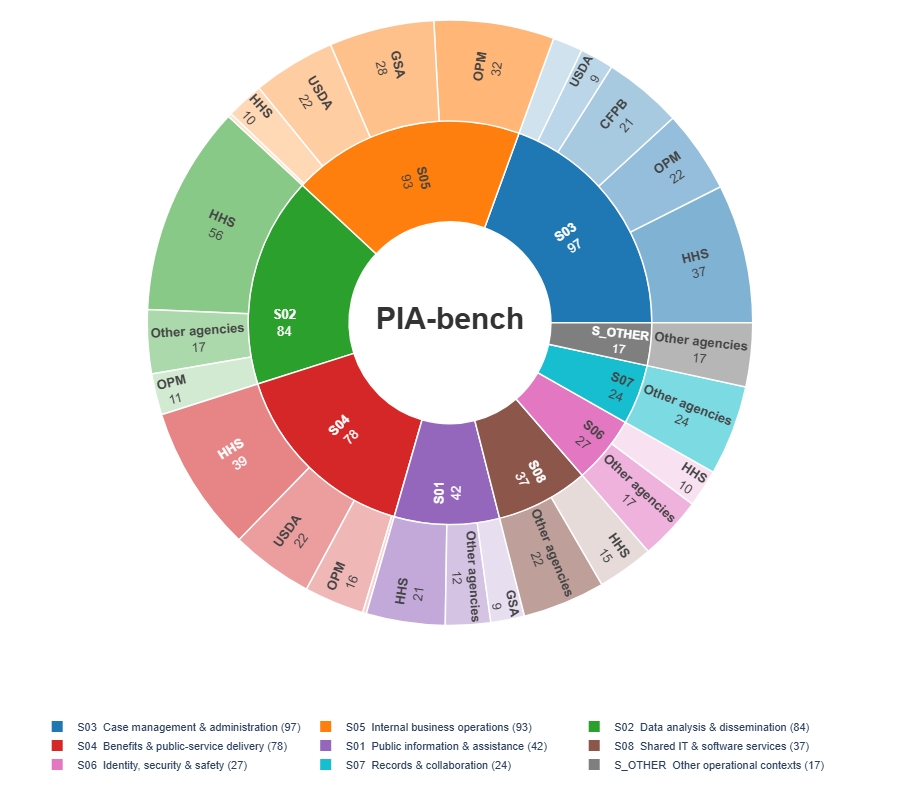}
    \caption{Prevalence of topical categories of PIA application scenarios and associated agencies in \dsname.}
    \label{fig:corpus_topics}
\end{figure}

\paragraph{Data collection.} We started with collecting a corpus of 499 PIA documents from the public repositories of five U.S. federal agencies: the Consumer Financial Protection Bureau (CFPB), the U.S. Department of Agriculture (USDA), the U.S. Office of Personnel Management (OPM), the U.S. General Services Administration (GSA), and the U.S. Department of Health and Human Services (HHS)~\citep{cfpb-pia, usda-pia, opm-pia, gsa-pia, hhs-pia}. PIAs from these federal agencies are published under a statutory obligation for public audit with a high quality standard~\citep{omb2016a130}. These PIAs documented expert assessment of digital systems spanning different sectors and application domains; full breakdown in Table~\ref{tab:system-contexts}, with topical prevalence by agency summarized in Figure~\ref{fig:corpus_topics}. PIA methodologies also vary across jurisdictions and authors, from lightweight questionnaires to comprehensive documentation~\citep{iwaya2024}. This heterogeneity motivates our subsequent curation process for LLM benchmarking.

\paragraph{Screening document structure.} We parsed the 499 PDFs into structured text in Python~\citep{auer2024docling}. We then screened the corpus by examining whether a PIA's privacy analysis could be easily isolated from its descriptive system context, avoiding leakage and priming in evaluation. We used GPT-5.4 as a classifier to examine each document's section headings and excerpts for this task. A PIA is labelled \emph{clean} when its privacy analysis can be separated from a coherent system description; documents are \emph{mixed} when the two are interleaved, and \emph{messy} when the privacy analysis cannot be reliably isolated by section. This produced 73 clean, 65 mixed, and 361 messy PIA records (Table~\ref{tab:corpusstats}). In this work, we evaluated using the 73 clean PIAs. 

\paragraph{Input context standardization.} After structural screening, we grouped adjacent text blocks in the 73 retained documents into 3,934 text sections based on their section or subsection headings (Table~\ref{tab:corpusstats})~\citep{auer2024docling}. Each text section preserves the original PIA section structure and covers a coherent topic, such as an assessment question and its response. To standardize system context, we use GPT-5.4 as a classifier to organize the unchanged text of each text section according to all applicable attributes in a set of eight attributes derived from established PIA frameworks: context, purpose, legal basis, information types, operations, recipients, retention, and assets (Table~\ref{tab:attrsources}).

\paragraph{Ground truth.} A separate LLM pass using the same model identifies verbatim passages within these sections, with each passage containing one or more consecutive sentences expressing a single privacy risk, proposed mitigation, or explicit statement that no risk was identified, following the schema in Appendix~\ref{app:gold}. These passages form the ground truth for \dsname{}. Across the 73 PIAs, we extract 451 passages for privacy risks, 831 for mitigations, and 95 explicit statements that no risk was identified. Validation against human annotations of 151 text sections shows 96\% recall and 100\% precision for passage identification (Appendix~\ref{app:validation}).

\section{Experiment and Analysis}
\label{sec:experiment_analysis}
We conducted exploratory experiments to demonstrate the potential of \dsname{} in benchmarking LLMs for real-world PIA.

\paragraph{Evaluation setup.} Given the standardized attributes for one PIA as input, we prompted each LLM to generate a structured JSON response with separate ranked lists of up to $K=10$ privacy risks or mitigations, each with a description, rationale, and supporting evidence quoted from the input. This cutoff is close to the pilot medians of nine risks and eight mitigations per PIA. Appendix~\ref{app:model_prompts} provides the complete prompts, response schema, and generation settings, including chain-of-thought (CoT) instructions. An LLM-as-judge (GPT-5.5) determines whether each generated description matches a ground-truth item under the same reference topic. We report precision, recall, and F1 at $k \in \{3, 5, 10\}$, macro-averaged across documents and three seeds, for five LLMs: four open models (Qwen2.5-7B~\citep{yang2024qwen25}, Qwen2.5-14B~\citep{yang2024qwen25}, Llama-3.1-8B~\citep{grattafiori2024llama3}, and DeepSeek-R1-Distill-14B~\citep{guo2025deepseekr1}) and one proprietary model (GPT-5.4 mini~\citep{openai2026gpt54mini}). We additionally report Flesch Reading Ease (FRE) scores~\citep{flesch1948new} to compare the readability of LLM-generated and expert-authored assessments.

\begin{table}[t]
\centering
\small
\setlength{\tabcolsep}{3.5pt}
\caption{Precision (P), recall (R), and F1 at $k{=}10$ against the agency's documented privacy risks and mitigations, macro-averaged over 73 documents and then over three seeds, alongside Flesch Reading Ease (FRE)~\citep{flesch1948new}. The random baseline scores each document against a different, randomly chosen document's real analysis, so it reflects generic privacy vocabulary shared across PIAs. The agency reference reports the documented items' own FRE. Best result per column in bold.}
\label{tab:pilot}
\begin{tabular}{@{}llccccccccc@{}}
\toprule
& & \multicolumn{4}{c}{Privacy Risk} & \multicolumn{4}{c}{Risk Mitigation} \\
\cmidrule(lr){3-6}\cmidrule(lr){7-10}
Model & CoT? & P & R & F1 & FRE & P & R & F1 & FRE \\
\midrule
Qwen2.5-7B & N & 0.39 & 0.17 & 0.21 & 19.7 & 0.50 & 0.25 & 0.30 & 18.7 \\
 & Y & 0.46 & 0.26 & 0.29 & 19.9 & 0.52 & 0.31 & 0.33 & 17.5 \\
Qwen2.5-14B & N & 0.45 & 0.28 & 0.32 & 15.6 & 0.47 & 0.36 & 0.36 & 14.9 \\
 & Y & \textbf{0.53} & 0.27 & 0.32 & 14.8 & \textbf{0.53} & 0.32 & 0.35 & 13.0 \\
Llama-3.1-8B & N & 0.29 & 0.39 & 0.30 & 23.0 & 0.35 & 0.49 & 0.36 & 11.0 \\
 & Y & 0.35 & \textbf{0.47} & 0.37 & 18.3 & 0.35 & \textbf{0.51} & 0.37 & 12.0 \\
DeepSeek-R1-Distill-14B & N & 0.33 & 0.36 & 0.32 & 25.2 & 0.35 & 0.45 & 0.35 & 14.1 \\
 & Y & 0.40 & 0.31 & 0.31 & 28.7 & 0.41 & 0.39 & 0.35 & 15.2 \\
GPT-5.4-mini & N & 0.38 & 0.40 & 0.36 & 18.5 & 0.42 & 0.49 & \textbf{0.42} & 18.1 \\
 & Y & 0.44 & 0.42 & \textbf{0.40} & 19.7 & 0.45 & 0.47 & \textbf{0.42} & 19.8 \\
\midrule
Random baseline  & -- & 0.19 & 0.19 & 0.18 & -- & 0.21 & 0.22 & 0.19 & -- \\
Agency reference & -- & -- & -- & -- & 29.2 & -- & -- & -- & 22.5 \\
\bottomrule
\end{tabular}
\end{table}

\paragraph{Overall performance.} GPT-5.4-mini with CoT achieves the highest F1 for privacy risks and mitigations (0.397 and 0.423), while zero-shot Qwen2.5-7B achieves the lowest (0.215 and 0.298). All systems exceed the shuffled-PIA floors (0.187 and 0.193), but even the best F1 remains below 0.43. Scaling Qwen2.5 from 7B to 14B improves zero-shot F1 by 0.100 for risks and 0.066 for mitigations, but CoT narrows these gains to 0.028 and 0.019. CoT also has inconsistent effects across models, ranging from $-0.004$ to $+0.077$ for risks and $-0.013$ to $+0.034$ for mitigations, suggesting that stronger structural reasoning is needed for PIA workflows. Furthermore, the readability of LLM-generated assessments, measured by FRE is consistently lower than expert-authored PIAs. Nevertheless, even expert-authored PIAs have a FRE score below 30, which is considered \textit{``very difficult''} to read and mostly aligned with other legal or technical documents. Finally, risk identification remains harder than mitigation generation: at $k{=}10$, mitigation F1 exceeds risk F1 for each of the all models under both prompts, averaging 0.363 versus 0.321. This suggests that the performance difference is not explained by readability.

\begin{figure}[htbp]
    \centering
    \caption{Topic distributions of ground truth from PIA documents: privacy risks and risk mitigations (left) and model miss rates (right). A miss occurs when no model output matches the human-expert identified privacy risks or risk mitigation. The analysis includes 181 privacy risks and 206 risk mitigations assigned to a single topic, pooled across five models, two prompting conditions, and three seeds.}
    \label{fig:topic_distribution}
    \includegraphics[width=\linewidth]{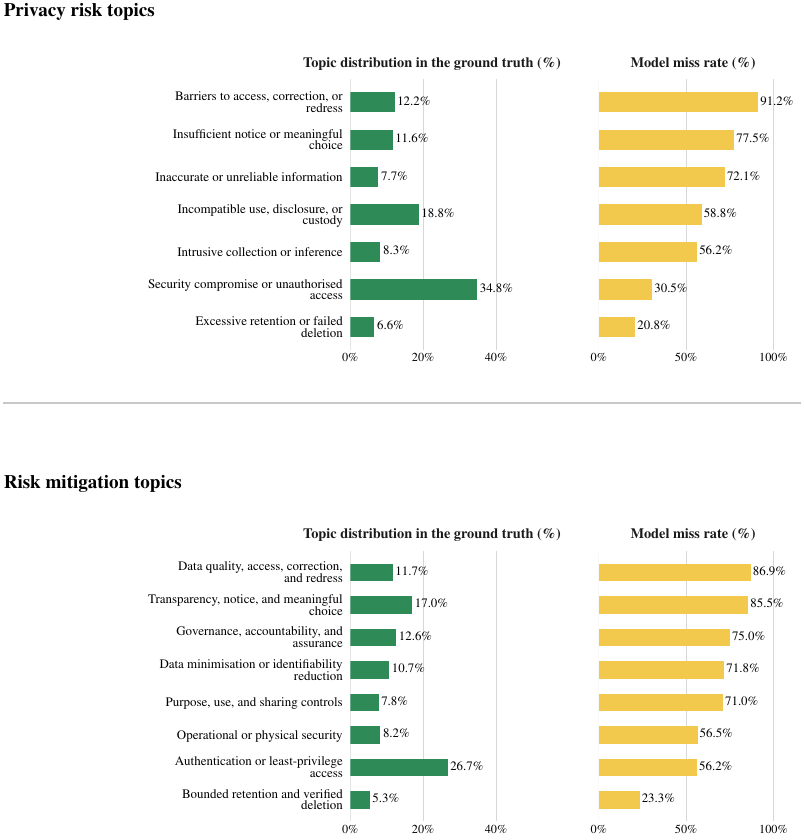}
\end{figure}

\paragraph{Error analysis.} Figure~\ref{fig:topic_distribution} compares topical prevalence with model miss rates for ground truth mapped unambiguously to a single named topic. Each ground truth miss rate is calculated over 30 runs per PIA (5 models $\times$ 2 prompts $\times$ 3 seeds). We assigned the topics using TopicGPT~\citep{pham-etal-2024-topicgpt}, an LLM-based topic modelling framework. Among topics of privacy risk, \emph{Barriers to Individual Access, Correction, Challenge, or Redress} has the highest miss rate (91.2\%). In contrast, \emph{(security compromise and unauthorised access)} is the most prevalent risk topic (34.8\%), which has a substantially lower miss rate (30.5\%). This difference suggests that models more reliably identify commonly defined core security and privacy issues than downstream or associated hazards that lead to privacy risks, such as information access, notice, and reliability issues.
Risk mitigation F1 is nevertheless at least as high as risk F1 in every condition. One possible explanation is that standard controls such as \emph{encryption and access restriction} recur across PIAs compared to risks that often depend on the practices of a particular system.

\paragraph{Implications for future work.}\label{sec:outlook} 

\textit{Integrating structural workflow and knowledge for LLM-powered PIA.} Our results point to future improvement of our proposed workflow in automating the identification of privacy risks and mitigations for PIA at scale with off-the-shelf LLMs. The inconsistent gains from model scaling and CoT further suggest that larger models or additional reasoning alone are insufficient to address this challenge. Our error analysis shows that performance varies across privacy issues: LLMs identify broadly defined security risks more reliably than risks concerning individual rights or system-specific mitigations. This motivates future work to explicitly incorporate the context and structure of PIA, including forms tailored to specific organizational workflows.
We encourage future work to develop dedicated tools and skills for LLM agents~\cite{schick2023toolformer} to adapt to the domain knowledge a PIA requires, building on established catalogs such as LINDDUN~\cite{deng2011linddun} and the CNIL knowledge bases~\cite{cnil2018pia}, as well as institutional resources. Beyond identifying privacy risks and proposing mitigation strategies, we envision LLM agents tackling other key steps in a PIA workflow, such as articulating and improving underlying system design decisions~\cite{danezis2014privacybydesign} or supporting privacy risk and control prioritization~\cite{nist2020privacyframework}.

\textit{Developing responsible and trustworthy LLM infrastructure for PIA.} Prior work highlights people's need to remain actively engaged in the PIA process in order to preserve agency and accountability~\citep{lee2026privy}, although an LLM itself could achieve satisfactory overall accuracy. Future work should therefore investigate different modes of human–LLM agent collaboration, including multi-agent teaming, to preserve human oversight, build trust, and integrate multiple viewpoints and critical thinking into PIA~\citep{turkstra2026argsbase}. Our exploration also motivates further improvements to the explainability and traceability of LLM output, which is not currently optimized to engage and support decision-making among stakeholders and developers whose domain expertise extends beyond legal, technical, and privacy knowledge. Moreover, existing software infrastructure could be redesigned to facilitate human–LLM co-working and improve collaboration efficiency with system designers, treating PIA as a live artifact to be maintained rather than a one-off documentation~\citep{tao2025privacy}.

\textit{Designing multi-dimensional assessment for PIA quality benchmarking.} In this work, we adopt standardized metrics (F1 and FRE) to assess LLMs' accuracy and readability in generating privacy assessments. However, metrics tailored to PIA quality are lacking in benchmarking, particularly with respect to downstream applications. PIA-specific assessment metrics would benefit the evaluation and improvement of both LLM-generated and human-authored PIAs. Future work can design and evaluate more aligned metrics from different perspectives, driven by both regulatory and practitioner requirements. For example, a potential metric could assess the specificity of proposed mitigations relative to the identified privacy risk and the system context~\citep{mollaeefar2025pillar}. Future work could also examine the consistency and robustness of LLM-proposed assessments when applied to a variety of new systems~\citep{bissoli2026linddun}. Additionally, one could evaluate the usefulness of proposed assessments and mitigations for system developers or privacy administrators, i.e., how readily the suggestions can be operationalized~\citep{lee2024practitioners}. 

\paragraph{Conclusion.} We introduce \dsname, the first benchmark based on public U.S. federal PIAs for evaluating whether LLMs can recognize privacy risks identified by human experts and propose mitigations from system context. Across five models, agreement with expert-authored assessments remains limited, CoT provides inconsistent gains, and models identify risks concerning individual rights less reliably than common security risks such as unauthorized access. While prior work sought to leverage LLMs for privacy assessment, our findings suggest that current LLMs are not yet suitable for automating formal PIAs. Such PIA assistance tools, instead, should combine PIA knowledge, structured workflows, evidence traceability, and human oversight. Future work should extend \dsname to less structured PIAs and other jurisdictions, involve practitioners in validation, and evaluate specificity, usefulness, and accountability.

\bibliographystyle{plainnat}
\bibliography{references}

\clearpage
\appendix

\section{Corpus Statistics}
\label{app:corpus}


\begin{table}[htbp]
\centering
\small
\caption{Structural class by agency over the 499 classified documents, with total source units among the 73 retained (clean) PIAs.}
\label{tab:corpusstats}
\begin{tabular}{@{}lcccccc@{}}
\toprule
Agency & Documents & Clean & Mixed & Messy & With a risk heading & Units \\
\midrule
CFPB  &  47 & 25 & 22 &   0 & 47 & 692 \\
USDA  &  87 & 45 &  5 &  37 & 40 & 3,140 \\
OPM   &  99 &  2 & 14 &  83 &  2 & 69 \\
GSA   &  66 &  1 & 24 &  41 &  0 & 33 \\
HHS   & 200 &  0 &  0 & 200 &  0 & -- \\
\midrule
Total & 499 & 73 & 65 & 361 & 89 & 3,934 \\
\bottomrule
\end{tabular}
\end{table}

\begin{table}[htbp]
  \centering
  \caption{Primary purposes of the systems and activities assessed in \dsname. Counts use one
  mutually exclusive primary-purpose label per PIA and sum to 73; 12 PIAs also span a secondary
  purpose. Definitions describe the dominant operational outcome, and examples are illustrative.}
  \label{tab:system-contexts}
  \begingroup
  \footnotesize
  \setlength{\tabcolsep}{3pt}
  \renewcommand{\arraystretch}{1.04}
  \begin{tabularx}{\linewidth}{@{}>{\raggedright\arraybackslash}p{0.29\linewidth}
                                  >{\raggedright\arraybackslash}X
                                  >{\raggedleft\arraybackslash}p{0.07\linewidth}@{}}
    \toprule
    \textbf{Primary purpose} & \textbf{Operational definition and examples} & \textbf{PIAs} \\
    \midrule
    Regulation and case administration
      & Manages formal matters arising from statutory or oversight duties, including regulatory reporting,
        supervision, enforcement, FOIA and Privacy Act requests, complaints, and appeals
      & 16 \\
    Public program and benefit delivery
      & Determines, administers, or delivers government benefits, loans, grants, insurance, or related
        services, including SNAP, farm loans, disaster assistance, and income support
      & 13 \\
    Research and public data
      & Produces general knowledge or data products rather than individual decisions, including research,
        public-use datasets, GIS, dashboards, reports, and analytical repositories
      & 11 \\
    Public communication and service channels
      & Informs, guides, or receives general requests and feedback from the public without initiating a
        formal case or benefit decision, through websites, social media, locators, contact centres, or surveys
      & 7 \\
    Internal administration
      & Supports the agency's own financial, workforce, facilities, procurement, or general operations,
        including billing, travel, payroll, timekeeping, and service requests
      & 7 \\
    Identity, physical security, and safety
      & Establishes identity or authorization, controls physical access, protects facilities, or monitors
        safety conditions, including credentialing, surveillance, and dam monitoring
      & 6 \\
    Records and collaboration
      & Creates, routes, stores, retrieves, images, retains, or shares records and files, including
        correspondence management, document imaging, and collaborative storage
      & 6 \\
    Shared technical infrastructure
      & Provides reusable hosting, integration, development, or security infrastructure for multiple
        applications, including Salesforce, common APIs, cloud platforms, and DevSecOps environments
      & 5 \\
    Other public operations
      & Covers clear purposes outside the preceding categories: federal fleet asset management and
        producer committee election administration
      & 2 \\
    \bottomrule
  \end{tabularx}
  \endgroup
\end{table}

\clearpage
\section{Gold Schema}
\label{app:gold}

Each gold span carries \texttt{span\_id}, \texttt{type} (\texttt{risk}, \texttt{mitigation} or
\texttt{null\_finding}), the verbatim \texttt{text}, the \texttt{unit\_ids} and
\texttt{sentence\_ids} it occupies, the \texttt{section} it was found in, a free-text
\texttt{explanation}, and a self-reported \texttt{confidence}. The sentence identifiers are what
allow the gold to be withheld at sentence rather than section granularity. Each link carries
\texttt{mitigation\_span\_id}, \texttt{risk\_span\_id} and a confidence.

\paragraph{Extraction prompt.} The system prompt is reproduced verbatim below, results are comparable only within a prompt version. The user message renders one
document chunk as numbered sentences, with the governing section heading or question inserted as a
marker line above the sentences it covers. Schema conformance is enforced at the tool-call layer and
every returned identifier is checked against the document before a span is accepted, so the model
selects content but cannot rewrite it.

\begin{promptbox}
You are annotating US federal Privacy Impact Assessments (PIAs) for a research benchmark.

You will be given the sentences of one PIA document, each with a numbered id. Your job is to find
the sentences where the agency states a PRIVACY RISK or a MITIGATION, and to link each mitigation
to the risk it addresses.

LABELS

risk
    A statement that privacy could be harmed, or that something undesirable could happen to
    personal information. Often hypothetical or conditional.
    Examples: "There is a risk that the PII collected may be used in ways that are not necessary
    for the purpose." / "Information being retained for an indefinite length can potentially be a
    risk."

mitigation
    A statement of what is done to reduce, prevent or manage a privacy risk. It must be offered as
    a response to a risk, not merely described as a feature of the system.
    Examples: "To mitigate this risk, the CFPB reviews collections of data within each application
    to minimize the data held." / "This risk is mitigated through the use of data archiving."

null_finding
    An explicit statement that no privacy risk exists for a topic.
    Example: "The CFPB's use of HUD data presents no privacy risks that relate to the purpose of
    the collection."

Sentences that are none of these are simply not returned.

BOUNDARY RULES, follow these carefully

1. A safeguard mentioned only as a property of the system is NOT a mitigation.
   "The system uses encryption at rest."                    -> not returned
   "To address this risk, the system encrypts data at rest." -> mitigation
   The test is whether the text presents it as answering a risk.

2. A question printed in the document is never a span. PIA forms contain questions such as
   "Are there any privacy risks for this system that relate to security?" — these are prompts, not
   findings. Only the agency's answer can be a span.

3. Do not treat routine security, training, auditing or access-control descriptions as mitigations
   unless the text connects them to a risk.

4. A sentence may belong to at most one span.

5. A span is one or more consecutive sentences that together make a single statement. Do not merge
   two separate risks into one span. Do not split one statement across two spans.

LINKING

After identifying spans, link each mitigation to the risk it addresses, by span_id. A mitigation
that addresses no stated risk is left unlinked. A risk may have several mitigations.

OUTPUT

Return sentence ids only. Never return sentence text. Give each span a short explanation of why it
qualifies, and a confidence of high, medium or low. Be conservative with "high".
\end{promptbox}


\section{Model Prompts}
\label{app:model_prompts}

\subsection{Zero-Shot User Prompt}

\begin{promptbox}
TASK. From the system context below, identify the privacy risks this system poses, and recommend a
mitigation for each risk, with a short justification grounded in the system.

SYSTEM CONTEXT
(*@\promptvar{SERVED\_INPUT}@*)

Instructions for a fair comparison.
- List up to (*@\promptvar{K}@*) distinct privacy risks, and up to (*@\promptvar{K}@*) distinct mitigations.
- Order both lists most important first. The order is the rank we score at each cut-off.
- Be comprehensive. Include every genuine, distinct risk you can support from the context, up to (*@\promptvar{K}@*).
- Do not pad. If the system genuinely has fewer than (*@\promptvar{K}@*) risks, list only the real ones and stop.
- Each item must be a single, self-contained risk or mitigation, not a bundle of several.
- If a mitigation is a standing control that does not answer any risk in your risks list, set
  "addresses" to an empty string.

Return JSON only, in this shape:
{
  "risks": [
    {"id": "r1", "description": "one distinct privacy risk in plain language",
     "evidence_quote": "verbatim span from the input that shows this risk", "rationale": "one sentence"}
  ],
  "mitigations": [
    {"id": "m1", "addresses": "r1", "description": "one recommended mitigation",
     "evidence_quote": "verbatim span, or empty string if you propose a new mitigation",
     "proposed": false, "rationale": "one sentence"}
  ]
}
\end{promptbox}

\subsection{CoT User Prompt}

The chain-of-thought condition states the same task and the same comparison instructions, inserts an
explicit five-step reasoning order after the system context, and adds a top-level
\texttt{reasoning} field to the response schema.

\begin{promptbox}
TASK. From the system context below, identify the privacy risks this system poses, and recommend a
mitigation for each risk, with a short justification grounded in the system.

SYSTEM CONTEXT
(*@\promptvar{SERVED\_INPUT}@*)

Reason step by step before you answer. Work through these in order:
1. What personal data does the system collect or hold, and who are the data subjects.
2. Who can access the data, and how does it flow, internally and to external recipients.
3. What could go wrong for the data subjects, given that data and those flows.
4. State each distinct privacy risk that follows.
5. For each risk, recommend a mitigation, preferring one the system already describes.

Instructions for a fair comparison.
- List up to (*@\promptvar{K}@*) distinct privacy risks, and up to (*@\promptvar{K}@*) distinct mitigations.
- Order both lists most important first. The order is the rank we score at each cut-off.
- Be comprehensive. Include every genuine, distinct risk you can support from the context, up to (*@\promptvar{K}@*).
- Do not pad. If the system genuinely has fewer than (*@\promptvar{K}@*) risks, list only the real ones and stop.
- Each item must be a single, self-contained risk or mitigation, not a bundle of several.
- If a mitigation is a standing control that does not answer any risk in your risks list, set
  "addresses" to an empty string.

Then return JSON only, in this shape:
{
  "reasoning": "2 to 4 sentence summary of the reasoning chain",
  "risks": [
    {"id": "r1", "description": "one distinct privacy risk in plain language",
     "evidence_quote": "verbatim span from the input that shows this risk", "rationale": "one sentence"}
  ],
  "mitigations": [
    {"id": "m1", "addresses": "r1", "description": "one recommended mitigation",
     "evidence_quote": "verbatim span, or empty string if you propose a new mitigation",
     "proposed": false, "rationale": "one sentence"}
  ]
}
\end{promptbox}

\subsection{LLM-as-Judge Prompt}
\label{app:judge_prompt}

Model used: \texttt{GPT-5.5}.

\begin{tcolorbox}[
    colback=gray!5,
    colframe=gray!50,
    boxrule=0.5pt,
    arc=0pt
]
\small
You judge whether predicted privacy items refer to the same underlying issue as the reference
items for one system. Two items match only if they concern the same privacy problem about the
same kind of data or data flow. Do NOT match two items just because both are privacy risks,
both mention security, or both use similar words. Be strict about specificity. When in doubt,
do not match.

\bigskip
\noindent\rule{\linewidth}{0.4pt}
\bigskip

\noindent
System task, \texttt{\{risks|mitigations\}} for one privacy assessment.\\[4pt]

REFERENCE \texttt{\{RISKS|MITIGATIONS\}} (the correct answers for this system):\\
\texttt{\{numbered\_reference\_list\}}\\[4pt]

PREDICTED \texttt{\{RISKS|MITIGATIONS\}} (to judge):\\
\texttt{\{numbered\_prediction\_list\}}\\[4pt]

Align predicted items to reference items. A predicted item matches a reference item only if it
names the same specific issue. Use each reference and each predicted item at most once, choose
the best one-to-one pairing.\\[4pt]

Return ONLY JSON: \texttt{\{"matches": [\{"predicted": <int>, "reference": <int>\}, ...]\}}.
If nothing matches, return an empty list.
\end{tcolorbox}

\Needspace{.4\textheight}
\section{Attribute Provenance}
\label{app:attrsources}

Table~\ref{tab:attrsources} lists each input attribute against the regulatory frameworks that
explicitly require it.

\begin{table}[htbp]
\centering
\small
\caption{Each input attribute against the frameworks that explicitly require it, with the document
wording it anchors to. F1 is OMB M-03-22 under Section 208 of the E-Government Act
\citep{egovact2002, omb2003}; F2 the minimum DPIA content of GDPR Article 35(7) \citep{gdpr2016};
F3 the Article 29 Working Party DPIA criteria \citep{wp29-2017}; F4 the CNIL PIA template
\citep{cnil2018pia}.}
\label{tab:attrsources}
\begin{tabular}{lll}
\toprule
Attribute & Framework support & Document wording anchors \\
\midrule
\texttt{context}           & F3, F4         & nature / scope / context \\
\texttt{purpose}           & F1, F2, F3, F4 & why collected / purposes / processing purposes \\
\texttt{legal\_basis}      & F3, F4         & lawfulness \\
\texttt{information\_types}& F1, F3, F4     & nature / personal data / data types \\
\texttt{operations}        & F2, F3, F4     & processing operations / functional description \\
\texttt{recipients}        & F1, F3, F4     & with whom shared / recipients \\
\texttt{retention}         & F3, F4         & period stored / storage duration \\
\texttt{assets}            & F3, F4         & assets on which data rely / data supporting assets \\
\bottomrule
\end{tabular}
\end{table}

\begin{promptbox}
You are annotating US federal Privacy Impact Assessments (PIAs) for a research benchmark.
You will be given the units of one PIA document. A source unit is the block of text under one heading covering a single subject, usually one question and its answer, sometimes several linked questions together, or a passage of narrative. Whatever its form, treat the unit as a single indivisible block of original document text for the purposes of this task.

For EVERY unit, identify which of the eight input attributes it explicitly provides.

THE EIGHT ATTRIBUTES
context             what the system/project is, its nature, scope and organisational setting
purpose             why the information is collected; the need, goal or intended benefit
legal_basis         the law, regulation, executive order or agreement permitting the processing
information_types   the categories and characteristics of personal information involved
operations          what is done with the information: obtained, checked, used, transformed,
                    analysed, transmitted, stored or deleted
recipients          people, roles, organisations or systems that receive it or have normal
                    operational access
retention           how long it is kept, or the event/schedule/rule determining disposal
assets              the technology and infrastructure supporting, hosting, storing or processing it

Use "none" when a unit provides none of the eight.

RULES
1. MULTI-LABEL. A unit often provides two or three attributes. Return all that apply. Do not
   force a single label.
2. The test is "does this unit EXPLICITLY provide this information", not "is it related to" and
   not "could a reader infer it".
3. Never combine "none" with a positive label.
4. Boundary cases that matter:
   - a goal without a system description is purpose, not context
   - a data source or collection method that does not identify the information is operations,
     not information_types
   - a recipient organisation is recipients, not information_types
   - technology named without an action is assets, not operations
   - storage with no duration or disposal rule is NOT retention
   - a data subject's right to access their own record is NOT recipients; operational access is
   - safeguards (encryption, authentication, monitoring, training) mentioned only as controls are
     none, unless the unit also states a recipient, data type or other positive attribute
5. The question text alone is not evidence. The agency's answer must provide the information.

OUTPUT
You are labelling, not rewriting. Do not extract, quote or paraphrase any text. For each unit you
tag with one or more attributes, the unit's own original text is retained as-is downstream as the
value for every attribute assigned to it; your only job is to decide which attributes apply.
Return one record per unit, using the unit_id exactly as given, with a confidence of high, medium
or low.
\end{promptbox}

\Needspace{.3\textheight}
\section{Validation}
\label{app:validation}


\begin{table}[htbp]
\centering
\small
\caption{Accuracy of identifying privacy risk and mitigation passages against independent human annotations of 151 text sections from three PIA documents. The structural baseline selects passages when risk or mitigation terms appear in a section heading or question, or when explicit labels appear in the text. }
\label{tab:main}
\begin{tabular}{@{}lcccc@{}}
\toprule
Method & Recall & Precision & F1 & Types and links passages? \\
\midrule
Language model                         & 0.960 & 1.000 & \textbf{0.980} & yes \\
Structural rules (heading + inline)    & 0.920 & 0.920 & 0.920 & no \\
\bottomrule
\end{tabular}
\end{table}

\clearpage
\section{Additional Experiment Results}
\label{app:result_diagram}

\begin{table}[htbp]
\centering
\small
\setlength{\tabcolsep}{4pt}
\caption{F1 for privacy risks and risk mitigations at ranked-list cutoffs
$k\in\{3,5,10\}$, macro-averaged over 73 documents and then over three seeds.
Best model result at each cutoff is bold.}
\label{tab:results_by_k}
\begin{tabular}{@{}llcccccc@{}}
\toprule
& & \multicolumn{3}{c}{Privacy Risks}
& \multicolumn{3}{c}{Risk Mitigations} \\
\cmidrule(lr){3-5}\cmidrule(lr){6-8}
Model & CoT? & $k{=}3$ & $k{=}5$ & $k{=}10$
             & $k{=}3$ & $k{=}5$ & $k{=}10$ \\
\midrule
Qwen2.5-7B
  & N & 0.205 & 0.215 & 0.215 & 0.289 & 0.298 & 0.299 \\
  & Y & 0.252 & 0.288 & 0.292 & 0.298 & 0.331 & 0.333 \\
Qwen2.5-14B
  & N & 0.271 & 0.311 & 0.315 & 0.314 & 0.359 & 0.364 \\
  & Y & \textbf{0.294} & 0.320 & 0.320
      & \textbf{0.319} & 0.351 & 0.351 \\
Llama-3.1-8B
  & N & 0.186 & 0.255 & 0.305 & 0.253 & 0.322 & 0.364 \\
  & Y & 0.252 & 0.329 & 0.372 & 0.269 & 0.341 & 0.373 \\
DeepSeek-R1-Distill-Qwen-14B
  & N & 0.230 & 0.289 & 0.319 & 0.242 & 0.312 & 0.354 \\
  & Y & 0.232 & 0.300 & 0.315 & 0.263 & 0.329 & 0.350 \\
GPT-5.4 Mini
  & N & 0.246 & 0.307 & 0.364 & 0.290 & 0.357 & 0.418 \\
  & Y & 0.269 & \textbf{0.350} & \textbf{0.397}
      & 0.309 & \textbf{0.385} & \textbf{0.423} \\
\midrule
Cross-document control
  & -- & 0.094 & 0.131 & 0.184 & 0.122 & 0.150 & 0.193 \\
\bottomrule
\end{tabular}
\end{table}


\end{document}